\documentclass[journal]{IEEEtran}
\usepackage{subfigure}
\usepackage{cite}
\usepackage{citesort}
\usepackage{amsmath,epsfig,bm}
\usepackage{pifont}
\usepackage{paralist}
\usepackage{graphicx}
\usepackage{amssymb}
\usepackage{amsmath}
\usepackage{cancel}
\usepackage[normalem]{ulem}
\usepackage[dvips]{color}
\usepackage{algorithm}
\usepackage{etoolbox}
\patchcmd{\thebibliography}{\section*{\refname}}{}{}{}

\usepackage{pifont}
\usepackage{graphicx}
\usepackage{amssymb}
\usepackage{amsmath}
\usepackage{cancel}
\usepackage[normalem]{ulem}
\usepackage[dvips]{color}
\usepackage{algorithm}
\usepackage{algorithmic}

\newcommand{\Hnull}{\mathcal{H}_0}
\newcommand{\Halt}{\mathcal{H}_1}

\newcommand{\Honull}{\mathcal{{D}}_0}
\newcommand{\Hoalt}{\mathcal{{D}}_1}

\begin{document}

\title{The Quantum Car}

\author{Robert~Malaney$^*$
%\emph{Member IEEE}
       % <-this % stops a space
%\thanks{}% <-this % stops a space

\thanks{$^*$Robert Malaney is with the School of
Electrical Engineering  and Telecommunications at the University
of New South Wales, Sydney, NSW 2052, Australia. email:
r.malaney@unsw.edu.au} }

%\author{
%\IEEEauthorblockN{ Robert Malaney}
%\IEEEauthorblockA{School of Electrical Engineering  \& Telecommunications,\\
%The University of New South Wales,\\
%Sydney, NSW 2052, Australia\\
%r.malaney@unsw.edu.au}
%}

\vspace{-3cm}

\maketitle
\begin{abstract}

%I explore the use of quantum information as a security enabler for the future driverless vehicle. Specifically, I investigate the role classical and quantum information can have on the most important  characteristic of the driverless vehicle paradigm - the vehicle location. Using  information-theoretic  verification frameworks, coupled with emerging quantum-based location-verification procedures, I show how  vehicle locations can be verified in a  manner simply not possible in classical vehicular networks. I discuss how quantum cryptography can be embedded directly in the location verification protocols, and how the same quantum states enabling the verification can lead to vehicle locations  being determined to arbitrary accuracy. The two technology enablers required for the quantum vehicular network are an increase in current quantum memory timescales and wide-scale deployment of the vehicular  communication infrastructure.  Current trends indicate these enablers should be in place well before the nominal ``20 year" deployment timescale normally attributed to emerging communication  technologies.

I explore the use of quantum information as a security enabler for the future driverless vehicle. Specifically, I investigate the role combined classical and quantum information can have on the most important  characteristic of the driverless vehicle paradigm - the vehicle location. By using  information-theoretic  verification frameworks, coupled with emerging quantum-based location-verification procedures, I show how vehicle positions can be authenticated with a probability of error simply not attainable in classical-only networks.
 I also discuss how other quantum applications can be seamlessly encapsulated within the same vehicular communication infrastructure required for location verification. The two technology enablers required for the driverless quantum vehicle are an increase in current quantum memory timescales (likely) and wide-scale deployment of classical vehicular communication infrastructure (underway). I argue  the enhanced safety features delivered by  the `Quantum Car'  mean its eventual deployment is inevitable.

\end{abstract}

%\section{Introduction}

 %Location information plays an ever-increasing role in modern telecommunication systems.

 \emph{Introduction -}  In next-generation  wireless networks, location information will be elevated to an almost pivotal role.
  %For example, in the 5G networks currently under investigation, it is expected that the very-high data throughput touted  will be delivered in part through the use of the location-based narrow-beamforming solutions of high-frequency antenna-array systems, e.g. \cite{5ga,5gb}.
%  %coupled to advanced non-linear tracking algorithms embedded in the antenna array's processer \cite{s7}. Without accurate location information 5G systems will be simply be unable to meet its advertised specifications.
 %Another example (and one in which 5G will eventually be deployed),
 %of an emerging wireless network in which location will play a pivotal role
 A clear example of this
 is in the emerging vehicular network paradigm, e.g., \cite{veh1,veh2}. In this paradigm, vehicles will communicate various attributes on themselves (e.g. speed, direction, location)   every 100~milliseconds. Such information will be combined with information gathered from on-board sensors such as laser systems, optical-recognition systems, and spatial-map processors, so as to deliver  the  driverless-car paradigm, e.g. \cite{driver1}. Great strides are already being made in this endeavor,
  %by most vehicle manufacturers, prodded along not only by consumer demand for safer vehicles, but by pending government regulation.
and many  would argue that  the (classical) driverless vehicle 1.0 has already arrived.
 One aim of this letter is to encourage  consideration of the enhanced benefits offered by the  driverless vehicle 2.0 - the quantum version.

  Clearly, the accuracy and verification of the location information in  vehicular networks will be a mission-critical requirement. However, upon trying to unconditionally verify the location of a vehicle solely within the classical realm we are immediately confronted by a noteworthy problem - it is impossible. Classical information can be copied, thus rendering any location verification system solely based on it attackable from a well-resourced adversary, e.g. \cite{chiang,ucla}.
  A potential solution to this is quantum location verification (QLV). As the name implies, QLV in essence adds quantum information into the location verification process.
  Such information, unlike its classical counterpart, cannot in general be copied exactly - the no cloning theorem \cite{woot}. This theorem leads to QLV
  %(which also underpins quantum cryptography)
  when it is coupled to another law of physics  - the no-signalling principle of relativity.

  The notion of QLV first appeared in the scientific literature in 2010, as a means of securing  real-time classical communications to a \emph{unique} spatial position \cite{mal1}.\footnote{Two months after \cite{mal1}  another  proposal on QLV appeared in the literature \cite{chan2}. Three months after \cite{chan2},  a 2006 patent  \cite{kent1}  on `tagging systems' for line-of-sight objects, based on entanglement, was first brought to my attention \cite{kent2}. To the best of my knowledge,  the patent \cite{kent1} first introduced the application of quantum information to object location verification.}   Since then many works related to QLV have appeared in the literature, e.g. \cite{mal2,chan2,kent3,q1,q2,q3,q4,q5}, largely focusing on  information-theoretic security issues under different conditions.
  As I discuss more later, it is widely accepted that under known attacks, QLV is \emph{effectively} secure. It is the main purpose of this letter to investigate the performance of QLV under the two most feasible attacks, showing how quickly it can overcome such  attacks to any required accuracy. I also highlight how other quantum technologies using the same QLV infrastructure lead to a much-enhanced,  and effectively \emph{unhackable},  vehicular network.

 %Irrespective of some deeper information-theoretic issues regarding whether QLV is formally unconditional (see later), there is little doubt in the community that the no-cloning and no-signalling principles can collectively deliver location verification schemes that are to all (pragmatic) intents and purposes, secure. The use of such QLV protocols in the context vehicular networks forms the core of this work.

 %\section{QLV in Vehicular Networks}
 %A schematic of a generic QLV is shown in Fig.~\ref{fig:1}.
\emph{QLV in Vehicular Networks -}  Within our system a reference station (RS)  is defined as any communicable device  within the vehicular network infrastructure  whose location  has already been authenticated.
 The prover vehicle (PV) in the  system is a vehicle which is yet to be authenticated (i.e. its claimed location is yet to be verified). I will assume that most (if not all) of the quantum information to be used in the QLV protocol is pre-stored  in the PV's on-board quantum memory. This information could have been delivered to the PV through an optical communication network (for example) when the PV was last hooked into its `electric-quantum' charging station.

 To help keep our adversary (Eve) on her toes, I will assume that the stored quantum states are non-orthogonal and take any allowable (possibly hidden) form, e.g. continuous variable (CV) states, qubits, qudits, hybrids, etc. I allow for some states being entangled  (possibly with states at multiple RSs).
  %We further keep Eve on her toes, by randomly selecting which combination of RSs will initially communicate with the PV at the onset of the QLV protocol.
   I will assume that coded signals sent from the RSs to the PV (instructing the PV in each round as to which operations to undertake   on which  quantum states) are wireless signals traveling at light-speed  - I refer to these signals as the verification information (VI). All VI (in a single round of the protocol) is obtained by the PV at the same instant.
   I will assume that the VI is encoded  across a subset of RSs (randomly selected at each round), and all communications between RSs are secure (e.g via quantum keys).
 % in general multiple inputs from multiple RSs are required for decoding.
  The actual number of RSs encoding the VI can also form part of the coded VI (i.e. a null signal from an RS can form part of the code). Coded VI received could imply instructions related to subsequent VI arriving $\epsilon_t$ time later. The number of bits of  information encoded in the VI signal (for at least some rounds) can be \emph{a priori}  unknown.  In general, the output of the PV upon receipt of the VI will be  classical and/or quantum signals sent to at least three RSs. To pass a verification test, the time taken to receive this output must be bounded by the round-trip time for the  RS-PV communications, plus the legitimate-receiver processing time.\footnote{Processing time is fundamentally bound by the energy available to evolve between orthogonal states (see \cite{bound}).
  An adversary must have  enough (undetectable) energy to ensure her additional combined processing time is undetectable. In the two feasible attacks studied here, I will assume enough energy is available to make  all processing delays effectively zero.}

 Our adversary, Eve, is formidable  - being fundamentally constrained only by the laws of physics. Most concerning for  our QLV system is that Eve has at her disposal an unlimited number of colluding devices.
 Eve's internal communications can be classical and/or quantum, and are limited only by the non-signalling principle. I assume Eve can transmit secretely and directly through anything, knows the locations of all RSs at all times, and can intercept any signal sent by any RS.
 %We also assume she has the ability to optimally clone all quantum information \cite{}.
  I also assume Eve's  \emph{finite} energy supply is limited only by the weak requirement that it is not detectable (e.g. via relativistic effects such as those influencing  GPS timings \cite{ash}).

   \emph{Gaussian States -}  For clarity, henceforth I will focus on the use of Gaussian CV quantum states (e.g. \cite{gauss} for review) within the QLV system. % Let us briefly introduce some of the formalism around such states (in the following we assume $\hbar=2$).
   In terms of the annihilation and creation operators $\hat a,\,{\hat a^\dag }$,  the quadrature operators $\hat q,\hat p$  defined for photon states are ($\hbar=2$ assumed)
$\hat q = \hat a + \,{\hat a^\dag }$ and $\hat p = i({\hat a^\dag } - \hat a\,)$,
  satisfying  $\left[ {\hat q,\,\hat p} \right] = 2i$.
 The quadrature operators for a CV state with $n$ modes can be defined by the vector
%\begin{eqnarray}\label{p2}
${\bm{\hat R}_{1, \ldots ,n}} = \left( {{{\hat q}_1},\,{{\hat p}_1}, \ldots ,{{\hat q}_n},{{\hat p}_n}\,} \right)$.
%\end{eqnarray}
%Similarly, ${R_{1, \ldots ,n}} = \left( {{q_1},\,{p_1}, \ldots ,{q_n},{p_n}\,} \right)$ is defined for the real variables $q, p$ - the eigenvalues of the quadrature variables.
 Gaussian states are characterized solely by the first moments  $\left\langle {{{ \bm{\hat R}}_{1, \ldots ,n}}} \right\rangle $ and a covariance matrix  $\bm{M}$, whose elements are given by
%\begin{eqnarray}\label{ad1}
$
{M_{ij}} = \frac{1}{2}\left\langle {{{\hat R}_i}{{\hat R}_j} + {{\hat R}_j}{{\hat R}_i}} \right\rangle  - \left\langle {{{\hat R}_i}} \right\rangle \left\langle {{{\hat R}_j}} \right\rangle$.
%\end{eqnarray}
%The covariance matrix of a $n$-mode quantum state is a $2n \times 2n$ real and symmetric matrix  satisfying the uncertainty principle.
  $\bm{M}$ can be transformed into  $\bm {M_s} = \left( {\begin{array}{*{20}{c}}
   \bm A & \bm C  \\
   {\bm{C}^T} & \bm{B}  \\
\end{array}} \right)$
%\begin{eqnarray}\label{p3}
%{M}= \left( {\begin{array}{*{20}{c}}
%A&C\\
%{{C^T}}&B
%\end{array}} \right),\,
%\end{eqnarray}
where
$\bm{A} = \tilde{a}{\bm{I_2}}\,,\,\bm{B} = \tilde{b}{\bm{I_2}}\,,\,\bm{C} = diag\left ( {{c_ + },{c_ - }} \right )$,
 $\tilde{a},\tilde{b},{c_ + },{c_ - } \in \mathbb{R}$, and
 ${\bm{I_2}}$ is the $2 \times 2$ identity matrix.
%The diagonal elements $a$ and $b$ are nonnegative real scalars and ${c_ + },{c_ - }$ are real scalars.
In this form the symplectic spectrum of the partially transposed covariance matrix is
$
{\nu _ \pm } = ( [ {{{\Delta  \pm \sqrt {{\Delta ^2} - 4\det \bm{M_s} } }} }]/ 2 )^{1/2}$ ,
%\end{eqnarray}
where $\Delta = \det \bm A + \det \bm B - 2\det \bm C$.
%A  measure of Gaussian entanglement can then be derived in terms of the logarithmic negativity ${E_{_{LN}}}\left( M_s  \right) = \max \left[ {0, - \log \left( {{\nu _ - }} \right)} \right]$ \cite{neg}. %where ${\nu _ - }$, as given above, is the smallest symplectic eigenvalue of the partially transposed covariance matrix.
From the symplectic spectrum many fundamental properties of Gaussian states can be derived (see \cite{gauss}).
An important Gaussian state  is the two-mode squeezed vacuum (TMSV) state,
%also known as the Einstein-Podolski-Rosen (EPR) state.
described for two modes $a$ and $b$ as
$\left| s \right\rangle  = \sqrt {1 - {\lambda ^2}} {\sum\limits_{n = 0}^\infty  {\left( { - \lambda } \right)} ^n}{\left| n \right\rangle _a}{\left| n \right\rangle _b},$ where $\lambda  = \tanh (r) \in \left[ {0,1} \right],$ and where
${\left| n \right\rangle _a}$ and ${\left| n \right\rangle _b}$ are Fock (number)  states of modes $a$ and $b$, respectively. Here,  $r$  is a   parameter quantifying the two-mode squeezing operator ${S_2}(r) = \exp \left[ {r\left( {\hat a\hat b - {{\hat a}^\dag }{{\hat b}^\dag }} \right)/2} \right]$.
%where ${\hat a}$ and ${\hat b}$ are the annihilation operators of the two modes.
The covariance matrix for the TMSV state can then be written
%can then be simply derived  in terms of  $v = \cosh (2r)$, the noise variance in the quadratures.
${\bm{M_{T}}} = \left( {\begin{array}{*{20}{c}}
   {v{\bf{I}}} & {\left( {\sqrt {{v^2} - 1} } \right){\bf{Z}}}  \\
   {\left( {\sqrt {{v^2} - 1} } \right){\bf{Z}}} & {v{\bf{I}}}  \\
\end{array}} \right)$
where the quadrature variance $v = \cosh (2r)$, and ${\bf{Z}}$~: = diag(1,-1).

 The TMSV state, being easily produced and manipulated, will likely play an important role in the
Quantum Car.
As well as being available for rounds of the QLV itself, the TMSV state could also be used  as part of a quantum key distribution (QKD) process, in which a vehicle which has passed a location verification test is then allowed to set up secret keys between itself and other RSs. Indeed, the infrastructure and resources required for QKD in the vehicular setting are very similar to those required by QLV. Further, some of the messages transferred during rounds of both protocols can be `double-dipped' upon. For example, classical information returned by the  PV  could also be used as part of the error estimation phase of QKD. Combined QLV/QKD can be achieved directly via the use of some entangled states shared by the PV and RSs  (or by the  logically equivalent `prepare and measure' schemes). Instruction on which states stored in memory are to be used for QKD can be sent as part of the QLV-VI request.
 %(for more detail on combined QLV/QKD see supplementary material \cite{supp}).
 % QKD protocols using the TMSV state as a basic resource are well known (e.g. \cite{weebrook}) and will not be described further here.

%\section{Decision Frameworks}

 \emph{Decision Frameworks -} Before discussing feasible threat models, let us first consider a generic formal decision framework based on some  observation vector $\bm{Y}$.  Our specific task within the context of the Quantum Car is to construct a  binary-decision framework for determining whether claimed location information (e.g. a GPS report) delivered in the IEEE~1609.2 frames  is to be trusted or not. If yes, the vehicle will retain (or be supplied with) a valid 1609.2 certificate; if no, the vehicle's existing certificates (if any) will be revoked (see \cite{veh2}).
 %As we shall see the notion of a ``trusted verification distance" (VTD) will enter the fray.
 The \emph{i}th element of the vector $\bm{Y}$ is of the form $Y_i=U_i+X_i$, with $U_i$ being the  value of a required metric estimated by (or from) an honest PV, and $X_i$ being some unwanted noise. For a malicious PV I assume $Y_i=V_i+X_i$, with $V_i$ being the value of a required metric being estimated by (or from) a malicious PV.   In all models considered here I  take $X_i$ to be a zero-mean normal random variable with  variance $\sigma_i$. Let $i=1\dots N$, $N$ being the number of RSs participating in the measurement process at a given round (in some circumstances fewer RSs can be compensated for by more rounds). For simplicity, I assume $\sigma_i=\sigma$ for all $i$.

 Let us consider two hypothesis, the null hypotheses $\Hnull$, and the alternate hypothesis $\Halt$. Under $\Hnull$ I assume the PV is legitimate, is at its claimed location (known by it exactly), and follows all coded instructions sent to it by the RSs in an honest and optimal fashion. Under $\Halt$ I  assume the PV is malicious, holds  as many devices as there are RSs participating in the measurement process,  with none of those devices actually at the claimed location. Assuming
 the observations collected by different RSs are independent,  under $\Hnull$ the  observation vector $\bm{Y} = [Y_1, \dots, Y_N]^T$ follows a multivariate normal distribution, $\bm{Y}|\Hnull \sim \mathcal{N} (\bm{U}, \bm{\Sigma}),$
where $\bm{U} = [U_1, \dots, U_N]^T$ is a mean vector, and $\bm{\Sigma} = \sigma\bm{I}_N$. Under $\Halt$ we have $\bm{Y}|\Halt \sim \mathcal{N} (\bm{V}, \bm{\Sigma}),$
where $\bm{V} = [V_1, \dots, V_N]^T$ is a mean vector (note we have implicitly assumed the variance is the same under both hypothesis).

Assuming a likelihood ratio test, which is known to be optimal for a wide range of cost metrics within the realm of classical location verification frameworks \cite{sh1,sh2,sh3},  the following decision rule for our system (written as a ratio of likelihood functions)  can be constructed,
%\begin{equation}\label{arbitrary}
$
\Lambda\left(\bm{Y}\right) \triangleq \frac{p\left(\bm{Y}|\Halt\right)}{p\left(\bm{Y}|\Hnull\right)} \begin{array}{c}
\overset{\Hoalt}{\geq} \\
\underset{\Honull}{<}
\end{array}%
\lambda,
$
%\end{equation}
 where $\lambda$ is the threshold, and $\Honull$ and $\Hoalt$ are the binary decisions that infer whether the prover is legitimate or malicious, respectively. Given our assumption of Gaussian noise this can be re-written as
%\begin{equation}\label{exp1}
%\Lambda\left(\bm{Y}\right) = \frac{\exp\left(-\frac{1}{2}(\bm{Y}-\bm{V})^T \bm{R}^{-1}(\bm{Y}-\bm{V}) \right)}{\exp\left(-\frac{1}{2}(\bm{Y}-\bm{U})^T \bm{R}^{-1}(\bm{Y}-\bm{U}) \right)}  \begin{array}{c}
%\overset{\Hoalt}{\geq} \\
%\underset{\Honull}{<}
%\end{array}%
%\lambda,
%\end{equation}
%Re-writing the above as
$\mathbb{T}(\bm{Y})
%\begin{array}{c}
\overset{\Hoalt}{\geq} \Gamma  $:
%\underset{\Honull}{<}
%\end{array}
$\mathbb{T}(\bm{Y})
%\begin{array}{c}
%\overset{\Hoalt}{\geq} \Gamma ,$
\underset{\Honull}{<}\Gamma ,$
%\end{array}
%\begin{equation}\label{decision_RSS}
%\mathbb{T}(\bm{Y})
%\begin{array}{c}
%\overset{\Hoalt}{\geq} \\
%\underset{\Honull}{<}
%\end{array}
%\Gamma,
%\end{equation}
where $\mathbb{T}(\bm{Y})$ is the test statistic given by
%\begin{equation}\label{statistic_RSS}
$
\mathbb{T}(\bm{Y}) \triangleq \left(\bm{V} - \bm{U}\right)^T \bm{\Sigma}^{-1} \bm{Y},$
%\end{equation}
and $\Gamma$ is a new threshold corresponding to $\mathbb{T}(\bm{Y})$ given by
%\begin{equation}\label{threshold_RSS}
$\Gamma \triangleq \ln \lambda + \frac{1}{2}\left(\bm{V} - \bm{U}\right)^T \bm{\Sigma}^{-1} \left(\bm{V} + \bm{U}\right).$
%\end{equation}
Note, in this notation, the false positive rate and detection rates are
%\begin{align}
$\alpha \triangleq \Pr\left(\mathbb{T}(\bm{Y}) \geq \Gamma |\Hnull \right), \
\beta \triangleq \Pr\left(\mathbb{T}(\bm{Y}) \geq \Gamma |\Halt \right), $
%\end{align}
respectively.
%\begin{equation}
%\alpha = \mathcal{Q}\left[\frac{\Gamma - \left(\bm{V} \!\!-\!\! \bm{U}\right)^T \bm{R}^{\!-\!1} \bm{U}}{\sqrt{\left(\bm{V} \!\!-\!\! \bm{U}\right)^T \bm{R}^{\!-\!1} \left(\bm{V} \!\!-\!\! \bm{U}\right)}}\right]\label{alpha_R}
%\end{equation}
%\begin{equation}
%\beta = \mathcal{Q}\left[\frac{\Gamma - \left(\bm{V} \!\!-\!\! \bm{U}\right)^T \bm{R}^{\!-\!1} \bm{V}}{\sqrt{\left(\bm{V} \!\!-\!\! \bm{U}\right)^T \bm{R}^{\!-\!1} \left(\bm{V} \!\!-\!\! \bm{U}\right)}}\right]\label{beta_R}
%\end{equation}
%respectively, where $\mathcal{Q}[x] = \frac{1}{\sqrt{2 \pi}}\int_x^{\infty} \exp (-t^2/2) dt$.
 Finally,  a cost metric is chosen, as this will  determine how to  set the threshold $\lambda$ (if the metric is optimized). A common choice for this is the total error  $ T_E=P_0\alpha +(1-P_0)(1-\beta)$, and I adopt that here with the \emph{a priori} probability of legitimacy set at $P_0=1/2$.
% \footnote{Note,
%this choice adopts a subjective determination on the costs of incorrect decisions. In order to avoid such a subjective framework  the normalized mutual information (NMI) could also be used as the cost metric \cite{yan2013optimal}.}
%*****************
Having built our decision framework, let us now investigate a specific use of our generic QLV system.
%The output of the PV, upon receipt of the VI, can take many forms. For example, the  PV could be simply  required to send a specific CV state (or states) stored in its memory to a specific RS (or RSs), or the output could  be simply  a classical output broadcast to all RSs.

In general, the PV does not know what operation(s) on what state(s) is needed at each round of the QLV until  the full VI for that round is obtained. Faced with this, Eve can take two quite feasible approaches to attack the system. One is to  wait until all the VI for that round is received, identify the CV state(s), determine which of her devices it is stored in, and then act on that state(s). The second is to try and copy the quantum information  optimally beforehand, distributing copies to all devices, and  act on those copies (as needed) as soon as the VI is fully received at each device.
We refer to the first strategy as the \emph{time-delay attack}, and the second strategy as the \emph{optimal-cloning attack}. Our  aim is to quantify the performance of the legitimate system against these two feasible attacks.
 %under a quadrature measurement (see \cite{gauss}).

\begin{figure}[!t]
    \begin{center}
  % {\includegraphics[width=3.5 in, height=2.5 in]{rob1n.eps}}
    {\includegraphics[width=4.0 in, height=1.2 in]{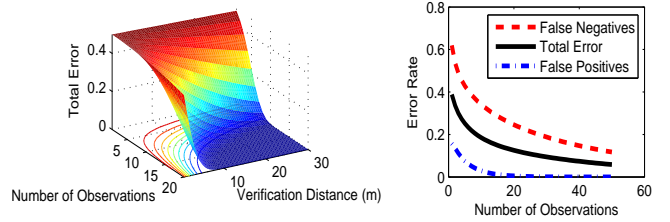}}
    \caption{(a) Total error ($T_E$) in QLV under a \emph{time-delay attack} (color online). Here the RSs are all randomly positioned, $\lambda=1$, and the standard deviation of the Gaussian noise in the timings is set at  $1 \mu$s. Eve has positioned her $N$ devices optimally,  each of them being some distance (the verification distance)  from a  claimed location. (b) $T_E$, the false negative rate  ($1-\beta$), and the false positive rate ($\alpha$) as a function of the  observation number (number of states tested), under an \emph{optimal-cloning attack}. Here 5 optimal copies from 1 state are used,  and $\lambda$ is set so that $\Gamma=2N$. Perfect quantum memory of the legitimate system is assumed, and timing information is assumed verified.  The attacks (a) and (b) are the most feasible (simplest) from a classical and quantum perspective, respectively. }\label{fig1}
    \end{center}
\end{figure}

Let us consider the time-delay attack where a quadrature measurement is requested in a given round on a specific stored CV state - the result of which is broadcasted.
%There are different $\Halt$ we could consider for our analysis under our first QLV scheme - the most pragmatic of which is a \emph{time-delay attack}.
 % Under this attack, the correct CV state is identified and measured,
   In the attack, the classical output is delivered by Eve's devices to the required RSs - albeit at some vector of delays ${{\bm{t}}_d}$ (due to the time needed for Eve to communicate the measurement outcome to all devices).
Using our previous relations we can determine the total error in the classification of the PV as a function of the verification distance and  the number of observations $N$.
%(note the number of rounds of the protocol can be encapsulated in a modified value of $N$).
 The results of such a calculation  are illustrated in Fig.~\ref{fig1}a (left), which shows  how the total error  can be made arbitrarily small by increases in $N$. Similar trends to Fig.~\ref{fig1}a can be found for other  operations on the states.
  %for any $N$ or verification distance.

In the optimal-cloning attack Eve utilizes a machine \cite{cloneo} that would  copy a CV quantum state to multiple (one for each RS) optimal-fidelity  clones\cite{clone1}.
 In such an attack there is no time delay incurred (relative to an honest PV). However,
 any operational measurement on the cloned state(s) will lead to different statistics relative to the original state(s).
   For example, consider when the VI requests some stored  CV states   be sent to specific RSs (the legitimate system has the states needed for any comparison test).
 Fig.~\ref{fig1}b (right) illustrates the case of a quadrature measurement operation used as a discriminating test
  in the scenario where the attack utilizes optimal clones  derived from a single coherent state.\footnote{The  discrimination error by this (readily implementable) technique is greater than the Helstrom bound \cite{hell}.
 %- the minimum  error probability in discriminating non-orthogonal quantum states.
 %In fact, for the cloning studied here I  find that the quadrature measurement test produces a weaker discriminatory test relative to  the upper limit $1/2\sqrt {F\left( {{\rho _0},{\rho _1}} \right)}$ \cite{fuchs}
%to the Helstrom bound  for two Gaussian states $\rho_0$ and $\rho_1$ ($F$ represents fidelity).
Quadrature measurements on other CV states (e.g. squeezed) can be optimal \cite{nha}.} Here I have used the general result that  for  $N_c\rightarrow M_c$  ($M_c \ge N_c$) cloning of $N_c$ general coherent states,  $M_c\ge 2$ being the number of equal fidelity optimal clones produced, the variance of the cloned states ${\sigma _{cl}}$
   will have an additional minimum variance $\left( {2/N_c - 2/M_c} \right){\sigma _0}$, where $\sigma_0$ is the variance of the vacuum noise \cite{clone1}. Discriminating between the different hypothesis in this circumstance takes a different form from a delay attack (the variance is no longer the same under both hypothesis). The test statistic for $N$ observations (now the number of  states tested)  becomes $\sum\limits_{i = 1}^N {{{({Y_{_i}} - {U_i})}^2}}$, and the threshold takes the form $\Gamma  = \left[ {\left( {2{\sigma _0}{\sigma _{cl}}} \right)/\left( {{\sigma _{cl}} - {\sigma _0}} \right)} \right]\left[ {\log \lambda  +N \log {{\left( {{\sigma _{cl}}/{\sigma _0}} \right)}^{0.5}}} \right]$ (here the $Y_i$'s and $U_i$'s now refer to quadratures).
 %Under the null hypotheses  $\sum\limits_{i = 1}^N {{{({Y_{_i}} - {U_i})}^2}/{\sigma _0}}$.
 Under these circumstances the false-positive rate is given by $1 - \chi _N^2\left( {\Gamma/\sigma_0 } \right)$, where $\chi _N^2(x)$ is the chi-square cumulative distribution
function with $N$ degrees of freedom at the value $x$. Similarly, the detection rate is given by $1-\chi _N^2\left( {\Gamma /\sigma_{cl} } \right)$. Fig.~\ref{fig1}b displays all the error rates for a  specific parameter setting. Again, we can see that reduction of the total error  with increasing $N$ is once again possible
 - and that the total error can in principle be made arbitrarily small.
 Similar trends to Fig.~\ref{fig1}b can be found for any cloning attack, including squeezed states and TMSV states (different tests may be used on the states).
%If TMSV states are used, such as in a combined QLV/QKD application, other discriminatory tests could be used (e.g.  correlations between the entangled modes).

 %Similar calculations to (b) can be similarly attained for squeezed states and TMSV states.

 Note, that I do not build into any $\Halt$ the possibility of a so-called teleportation attack \cite{q2}. Several issues can justify this. Arguably, the most important issue is a pragmatic one - the amount of resources needed for the attack in general appear unfeasible.\footnote{The best bound  on the number of entangled pairs needed for a successful teleportation attack (with unit fidelity obtained in the infinite  port-number limit \cite{port})  is of the form $2^{O(n)}$,
  where $n$ is number of states  to be teleported\cite{q3} (I remind the reader that $2^{270} \sim$  the number of nuclei in the observable universe). Pragmatism aside, an exponentially increasing amount of entanglement can eventually become detectable through its associated  energy. Creation of any bipartite entanglement costs energy  for temperatures $T>0$ \cite{huber}. Reaching $T=0$ requires infinite heat extraction.
  %It does appear wherever you look, general teleportation attacks formally require the use of some infinite resource.
 } In this work I simply adopt this view.
   However, I do also note, from an information-theoretic perspective, that in general the required teleportation  needed  for the attack is not available.\footnote{By this I mean with zero probability of detection using finite resources. That said,
    the attack  of \cite{q2} poses a formidable in-principle challenge to QLV. For example, in the  QLV of \cite{mal1}
$2^N$~$N$-bit messages are encoded in a non-orthogonal ensemble of stored entangled states, the ensemble being decoded later via classical inputs. Infinite resource issues aside, specific cases of \cite{mal1} (e.g.  dimension-$N$ maximally-entangled qubit states with no null-signalling code component) are in-principle attackable by the \emph{explicit} procedure given in \cite{q2}. Expanded expositions of the main concept in \cite{q2} are possible, but ultimately all energy requirements (e.g. teleportation, unitary evolution,  erasures) must  be bounded.}
   For example, in CV-based QLV schemes the required CV teleportation (deterministic with unit fidelity) is not possible at finite energy.\footnote{Of course, realistic energy levels can push the CV teleportation fidelity arbitrary close to 1, but a non-unit fidelity is in-principle detectable (multiple rounds of teleportation at finite energy should enhance such detection).} As such, it may be useful to look at the question posed by a teleportation attack from more of a \emph{limited} legitimate-system viewpoint rather than from an attacker's \emph{in-principle} viewpoint.\footnote{
   %Given its seemingly \emph{in-principle} requirement for some infinite resource, and given the fact the legitimate system can in-principle perform operations to the same degree as the adversary, perhaps the role of teleportation attacks is best posed in a more pragmatic manner.
    E.g. - For some $0<\epsilon<1$, under what restrictions on the operations  of a legitimate system, can an adversary using some defined finite resource produce an undetectable attack with a success probability $1-\epsilon$?} %($\epsilon>0)$?}

   %but \emph{realistic} levels of entanglement will in general not lead to a successful attack.  In any formal discussion, energy arguments  must be given the same \emph{in-principle} context afforded to any attack being proposed.}
%

 Finally, I discuss how some quantum states can influence timing measurements made  within the vehicular network.
 %Using only  classical wireless signals the accuracy of a location estimate is given by the classical Cramer-Rao lower bound (e.g \cite{loc1,loc2}). 
 Assuming classical only signals,
   and Gaussian noise with variance $\sigma_t$  in all the  timing measurements,   the variance $\nu_{cr}$ of the position error of a device located at $(x_0,y_0)$ satisfies,
 $\nu _{cr}^{1/2} \ge c\sqrt {{\sigma _t} N} {\left( {\sum\limits_{i = 1}^{N - 1} {\sum\limits_{j = i + 1}^N {{{\sin }^2}\left( {{\varphi _i} - {\varphi _j}} \right)} } } \right)^{-1/2}}$,
 where the \emph{i}th RS is at $(x_i,y_i)$,  $\tan {\varphi _i} = \left( {{y_i} - {y_0}} \right)/\left( {{x_i} - {x_0}} \right)$ and $c$ is the speed of light (e.g. \cite{loc1,loc2}).
 For some estimated position,  we can see this leads to a $1/\sqrt{N}$ dependence for the location error. Implicit in the above bound is also a $1/\sqrt{N}_p$ dependence on the mean number of photons, $N_p$, used in each timing measurement.
 %However, by utilizing \emph{quantum positioning}  this dependency can be turned into  $1/N$ \cite{loc3}. 
 Quantum positioning \cite{loc3} invokes the use of entangled states  to shift this dependence to the form
  $1/{N}$, with squeezed states being used to provide a dependence of the form $1/N_p$. Fundamentally, both  gains can be traced back to a higher spread in energy in the non-classical states, relative to coherent states. The resulting improved timings  could be used to enhance the verification-decision process, or be used
 %when the VI requests some CV entangled states be sent to specific RSs.
 to  simply enhance actual (or relative) position  estimates of vehicles.
 %the positioned derived directly as an improvement on any claimed location such as a reported
% GPS position (perhaps \emph{relative}) positioning scheme,
 %Most likely such enhanced positioning techniques would be used in the context of vehicle-to-vehicle (V2V) communications
  For example, consider positioning via the use of laser signalling between vehicles. In this scenario,
  location errors in the \emph{sub-}\emph{mm} range can be anticipated using levels of  entanglement and squeezing already achievable in the laboratory. Classical wireless positioning can sometimes lead us into the realm of the `dark arts' (e.g. removal of biases, nuisance parameters \cite{loc2}). The addition of quantum positioning techniques can only shed (non-classical) light.
 %Ultimately,
   %positioning is in practice more complex than that discussed here (detection and removal of biases, nuisance terms, etc, \cite{loc2}),
    %in the vehicular network environment using laser-based vehicle-to-vehicle communications, location verification distances of \emph{sub-mm} can be anticipated, with the accuracy fundamentally bounded by the  energy (spread) within the entangled states.

 \emph{Outlook -} Our driverless vehicle 2.0 is certainly of the future - but not the \emph{distant} future. The two main enablers needed - quantum memory of 1-day lifetime (or time between electric-quantum charges) and widespread  vehicular communication infrastructure  - appear  within reach. Advances in quantum memory lifetimes are improving dramatically, with the current record (at a temperature of 2K) at 6 hours \cite{memory2}. Confidence is high that photonic long-term quantum memory at \emph{car-boot} temperature is achievable in the coming years \cite{memory}.
  Aside from the initial delivery of the quantum information (e.g. via optical fibre), QLV `in-the-field' can operate with only classical  wireless communication infrastructure - deployment of which in the wider vehicular context has already commenced.
  Of course, the driverless vehicle 3.0 will be even further advanced,  deploying additional quantum technologies and applications.\footnote{For example, entanglement (perhaps macroscopic) shared between vehicles for enhanced inter-vehicle control and  network synchronization.} But 2.0 may keep our vehicle engineers busy, at least for now.

  %increasing by 3 orders of magnitude in the past few years - to a state-of-the art 1 second \cite{memory}. A similar increase in the next few years will see us where we need to be.

 \emph{Conclusions - }
 The quantum technologies discussed here will have many applications, offering unparalleled security and sensitivity in many  scenarios. However, if the \emph{number of lives saved} is the optimization metric of choice, use of these technologies in a widely-deployed vehicular network could well deliver the \emph{optimal} quantum-communication application.

 %(unconstrained entanglement and the ability to conduct an indefinite number quantum of operations at no time penalty).

%\subsection{Preliminaries}

%

%
%
%\appendix

\end{document}